# Integrating Power Electronics-based Energy Storages to Power Systems: A Review on Dynamic Modeling, Analysis, and Future Challenges

*Qiang Fu*[a,b], *Changlong Dai*[b], *Siqi Bu*[a,*], *C. Y. Chung*[a]

[a] *Department of Electrical and Electronic Engineering, The Hong Kong Polytechnic University, Hong Kong, 999077, China*
[b] *School of Electrical Engineering, Sichuan University, Chengdu, 610065, China*

***Abstract:*** The integration of power electronics-based energy storage systems (PEESs) into power systems introduces potential instabilities. This study reviews efforts in dynamic analysis of both AC and DC power systems integrated with PEESs, covering dynamic modeling, analysis methods, and potential instability risks. Major conclusions are drawn as: 1) Simplified models of PEESs have been widely used for dynamic analysis of power systems. However, it may cause "error aggregation" as the scale of PEESs increases, leading to mistakes in results, which induces significant concerns. 2) Traditional stability mechanism analysis methods remain effective for single grid-connected PEES and large-scale PEESs with parallel and series connections. However, they are inadequate for PEESs with distributed connections. To fill in this gap, an idea of mechanism analysis based on "dynamic reconstruction" is proposed. 3) Potential instability risks caused by PEESs integration may differ from those caused by renewable energy integration due to differences in functional controls and bidirectional power flow. However, comprehensive investigations in this regard are lacking and require significant attention. To ensure the stable operation of power systems with increasing integration of PEESs, significant challenges are summarized in the end, providing inspirations for future studies.

***Highlights:***

- The unique characteristics of power electronics-based energy storage systems are clarified.
- The impact of error aggregation in modeling is noted and analyzed.
- Dynamic reconfiguration is proposed to address high-dimensional stability issues.
- Challenges caused by bidirectional power flow and functional controls are identified.

**Word count: 8,947**

**List of abbreviations:**

| | |
|---|---|
| PEESs | Power electronics-based energy storage systems |
| REs | Renewable energies |
| ESs | Energy storages |
| ESSs | Energy storage systems |
| MEESs | Mechanical equipment-based energy storages |
| ESM | Energy storage module |
| GFL | Grid-following |
| GFM | Grid-forming |
| PCC | Point of common coupling |
| PLL | Phase-locked loop |
| EMS | Energy management systems |
| MIMO | Multi-input and Multi-output |
| GNC | Generalized Nyquist criterion |
| SISO | Single-input and Single-output |

* = corresponding author details, siqi.bu@polyu.edu.hk

## 1. Introduction

In recent years, rapid development in renewable energies (REs), such as photovoltaics and wind generators, has been propelled with low-carbon requirements in power systems [1], which is confirmed by the global policies related to REs. For example, the United States introduced the "Energy Policy Act of 2005" in 2005 [2], Germany revised the "Renewable Energy Sources Act" in 2017 [3], and the European Union proposed "REPowerEU" in 2022 [4], etc. As illustrated in Fig. 1, the developing trend of REs is depicted by green lines [5].

However, the random and volatile nature of output power from REs worsens the dynamic characteristics of power systems, raising concerns from studies. Some view it as unreliable energy, causing significant frequency variations [6], low inertia [7], long recovery times [8] in power systems, and potential negative impacts on the surrounding environment [9]. Consequently, certain studies attempt to address these issues by installing energy storages (ESs) [10], which have emerged as an effective solution to mitigate power volatility by actively balancing power as needed. Motived by which, the deployment of energy storage systems (ESSs) has experienced substantial growth in recent years, with projections indicating that by 2030, ES capacity will reach 1,867 GWh [11], making it a crucial component of future power systems. Due to delayed official data updates, the developing trend of ESs is illustrated up to 2023 with orange lines in Fig. 1 [12], and major projects are listed in Table 1.

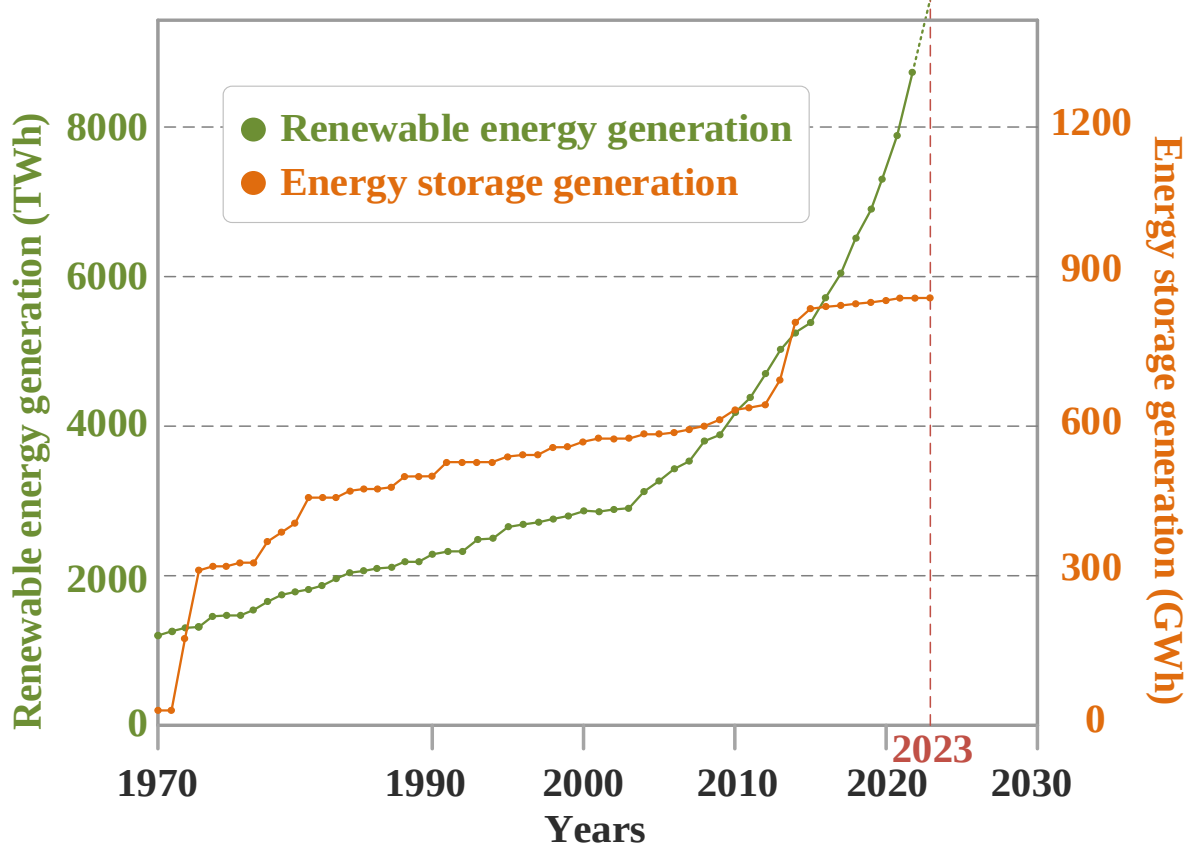

Fig 1. Developing trend of renewable energies and energy storages.

**Table 1**
List of energy storage projects

| Projects | ESM | Year | Location | Rated Power |
|---|---|---|---|---|
| Three Gorges Dam [13] | Pumped Hydro | 2003 | Hu Bei, China | 22500 MW |
| Audi E-Gas Project [14] | Hydrogen | 2013 | Werlte, Germany | 6 MW |
| Beacon Power Flywheel Energy Storage [15] | Flywheel | 2014 | New York, United States | 20 MW |
| Goderich Energy Storage Centre [16] | Compressed Air | 2019 | Goderich, Canada | 1.75 MW |
| FuelCell Energy Renewable Biogas Fuel Cell [17] | Fuel Cells | 2019 | California, United states | 2.8 MW |
| Moss Landing Vistra Battery [18] | Lithium-Ion Batteries | 2021 | California, United States | 400 MW |
| Energy Superhub Oxford [19] | Redox Flow Battery | 2020 | Oxford, United Kingdom | 2 MW |
| Manatee Energy Storage Center [20] | Batteries | 2021 | Florida, United States | 409 MW |
| Superconducting Magnetic Energy Storage Demonstration Unit [21] | Superconductor Magnetics | 2022 | Saint-Ouen, France | 20MW |
| Frequency Regulation Project [22] | Supercapacitor | 2023 | Fu Jian, China | 2.5MW |

There are two types of grid connections in ESs: power electronics-based energy storages (PEESs) and mechanical equipment-based energy storages (MEESs). As depicted by the green area in Fig. 2, MEESs convert different forms of energy into electricity through mechanical equipment between the energy storage module (ESM) and the connected power system using mechanical equipment such as synchronous generators, specially, flywheel ESs [23], pumped hydro ESs [24] and compressed air ESs [25]. In contrast, as depicted by the orange area in Fig. 2, PEESs convert different forms of energy into electricity using power electronics, such as DC/DC or DC/AC converters. These converters support bidirectional power flow. Therefore, in this study, the term "DC/AC converters" refers to devices that can convert both AC to DC and DC to AC, rather than functioning solely as inverters. Particularly, hydrogen ESs [26], Lithium-Ion battery ESs [27], fuel cell ESs [28], supercapacitor ESs [29] and superconducting magnetic ESs [30]. Notably, the same ESM may correspond to different types of ESs when the grid connection ways are different. For example, a thermal ES stores energy by increasing the temperature of materials, if this is realized on the basis of mechanical generators, it is classified as MEESs. Alternatively, if the energy transfer relies on converters, it is classified as PEESs.

The reason for classifying ESs based on grid connections is that the dynamics of ESs primarily depend on the grid-side equipment. There are some differences between MEESs and PEESs. MEESs typically respond slowly to system changes, with response times in the order of seconds and minutes [31], making them less able to quickly respond to fast variations caused by REs. PEESs can achieve much faster responses to variations in frequency and voltage [32–37], down to the millisecond range, thanks to the rapid control capabilities of power electronics [38]. Moreover, MEESs have higher rated power capacities but are limited to specific geographical locations, resembling a centralized ES [39]. PEESs achieve large power capacity by integrating a larger number of ESMs, which can be either centralized or distributed, providing more flexibility [39]. With the development of REs, PEESs have garnered significant attention due to their better adaptability.

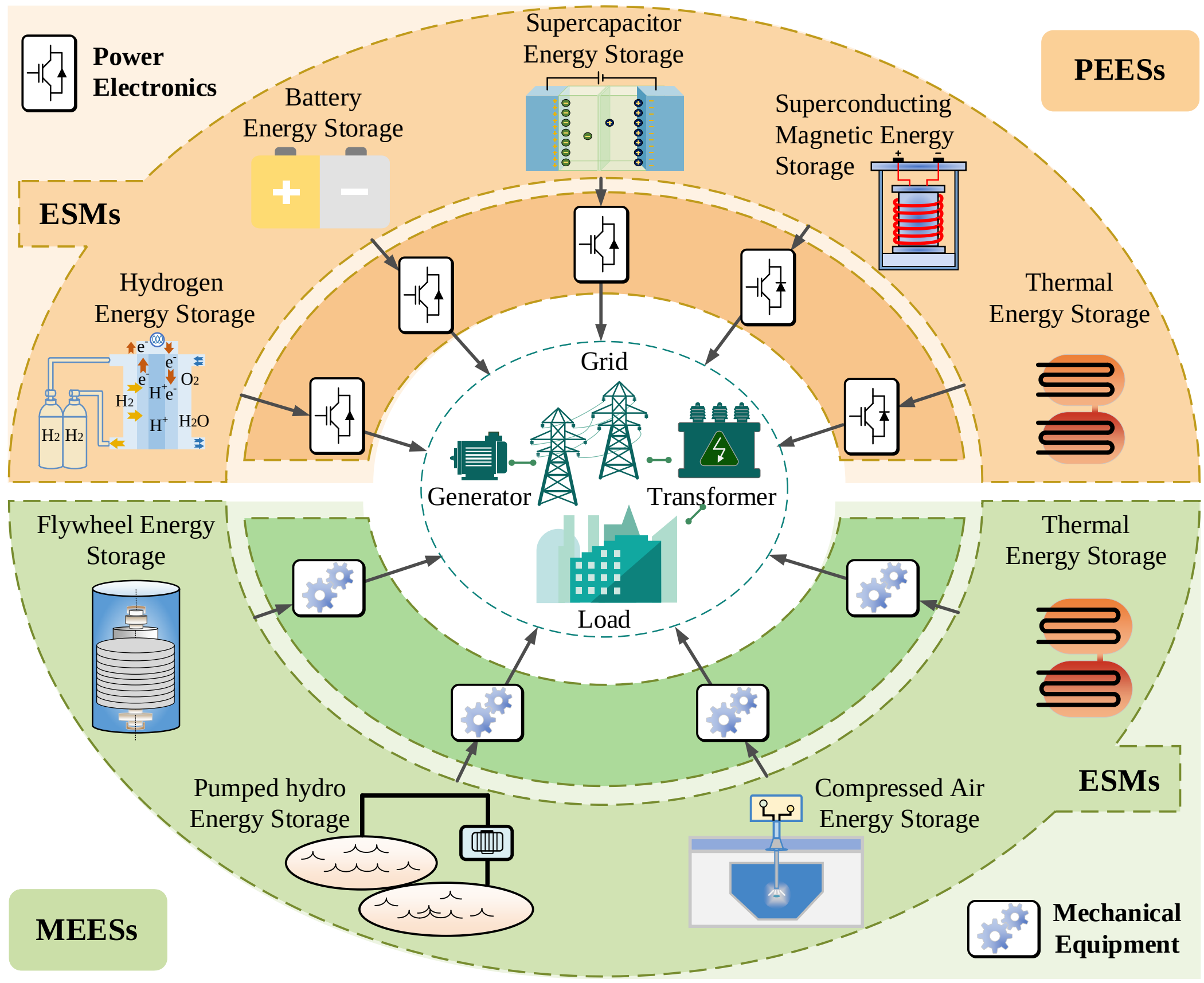

Fig 2. Illustration of PEESs and MEESs.

In addition to power balance [40], PEESs can offer active support to power systems due to their high controllability. A function of energy arbitrage was introduced in [41], where PEESs obtain electricity from the grid during low-demand periods and sell electricity to the grid during peak demand periods. PEESs were used for voltage improvements, such as reducing overvoltage during faults [42] and resolving voltage excursions in low-voltage distribution networks [43]. In [44,45], PEESs improved frequency dynamics. They contribute virtual inertia, mitigating the decline of the lowest frequency, and actively participate in primary frequency regulation by either absorbing or generating active power under significant disturbances.

However, integrating PEESs into power systems may also pose instability risks, especially in resonance instability and converter-driven instability, as recommended by the Power System Dynamic Performance Committee of IEEE [46]. These instabilities have been demonstrated in RE integration of under various conditions. However, such findings may not be entirely applicable to PEESs due to the following three inherent differences between PEESs and REs: Firstly, PEESs normally serve local grids instead of long-distance transmission, reducing the occurrence of weak connections, which is a major instability risk for REs. Secondly, PEESs offer a variety of functional controls for both DC and AC power systems, whereas controls of REs are limited due to the restricted power generation capacity. Thirdly, PEESs can charge or discharge power as need, the power flow is bidirectional, in contrast to REs which always output power to grids.

Therefore, apart from steady-state characteristic studies of PEESs [47–49], the dynamics of power systems integrated with PEESs need concerns, and thus clarifying potential instability risks of connecting PEESs, particularly in light of the rapid development of ESSs.

To comprehensively gather literature related to the stability of power systems integrated with PEESs, this paper first selected Web of Science, ScienceDirect, and IEEE Xplore as the primary databases for the literature search. Journal articles were then screened based on relevant keywords, resulting in a total of 3,331 papers. Of these, 1,327 papers were further filtered by limiting the publication years to the last decade and considering impact factors. Ultimately, 134 articles were retained for the review. Additionally, although some classic literature was published earlier, their theoretical value remains significant and justifies their citation. These papers are widely recognized not only for their frequent citations in academia but also for their authority and influence. Moreover, given that many key data points were not fully presented in research articles, a variety of information sources were consulted, including websites and professional reports. For example, data from globally authoritative databases such as the Department of Energy (DOE) and the Energy Institute (EI) were utilized to ensure the comprehensiveness and reliability of the results obtained.

In this study, the authors indicate that future integration of PEESs may suffer instability issues when the scale becomes large. To proactively address potential instability risks, studies on PEESs are reviewed, summarizing key findings that offer valuable insights for advanced control design to enhance dynamics of power systems integrated with PEESs. Major contributions are as follows:

1) Typical models for characterizing the dynamics of PEESs are summarized. When analyzing stability, the modeling of the dynamics of various types of ESMs and converters is often simplified to constant sources [50–52], which may introduce unacceptable levels of errors [53–55]. It is declared in this study that errors induced by simplified models will be magnified as the scale of PEESs expands, potentially leading to erroneous dynamic analysis results. Therefore, for future work, the necessity of numerical evaluation of model errors is emphasized, and an idea is proposed to help achieve this aim.
2) The instability risks of power systems integrated with PEESs using popular GFL and GFM controls are summarized, and major risk factors are identified. Considering the active support capability of PEESs, multiple types of functional controls have been proposed. However, the instability risks associated with these controls have not been comprehensively analyzed yet. Therefore, a thorough instability risk examination is encouraged to be conducted to mitigate potential risks posed by new controls.
3) Studies on the stability mechanism of power systems connected with PEESs are reviewed. It is concluded that traditional stability mechanism analysis methods are effective for single grid-connected PEESs and PEESs with series and parallel connections. However, elucidating the stability mechanism of PEESs with distributed connections remains challenging. Therefore, for future work, an idea of "dynamic reconfiguration" is proposed to assist in stability mechanism analysis, which can provide insights for developing novel stability mechanism analysis methods.
4) The differences between PEESs and REs are clarified, and the challenges of power systems connected with PEESs, distinct from REs, are summarized. It is evident that the impact of bidirectional power flow on power system dynamics becomes more pronounced as the scale of PEESs increases, leading to uncertainties in both source and load regions. However, current investigations into bidirectional power flow are insufficient. Therefore, for future work, the impact of bidirectional power flow on the stability of power systems connected with PEESs needs to be comprehensively analyzed.

The remainder of this study is organized as follows: In Section II, fundamental constructions and control strategies of PEESs are introduced, and the differences from REs are clarified. In Section III, major dynamic models of PEESs are summarized, and the modeling errors are analyzed. In Sections IV and V, studies on dynamics of DC and AC power systems integrated with PEESs are reviewed, respectively. Finally, Section VI draws the conclusions and discusses future challenges.

## 2. Construction and Control Strategies of PEESs

### 2.1. *Construction of PEESs*

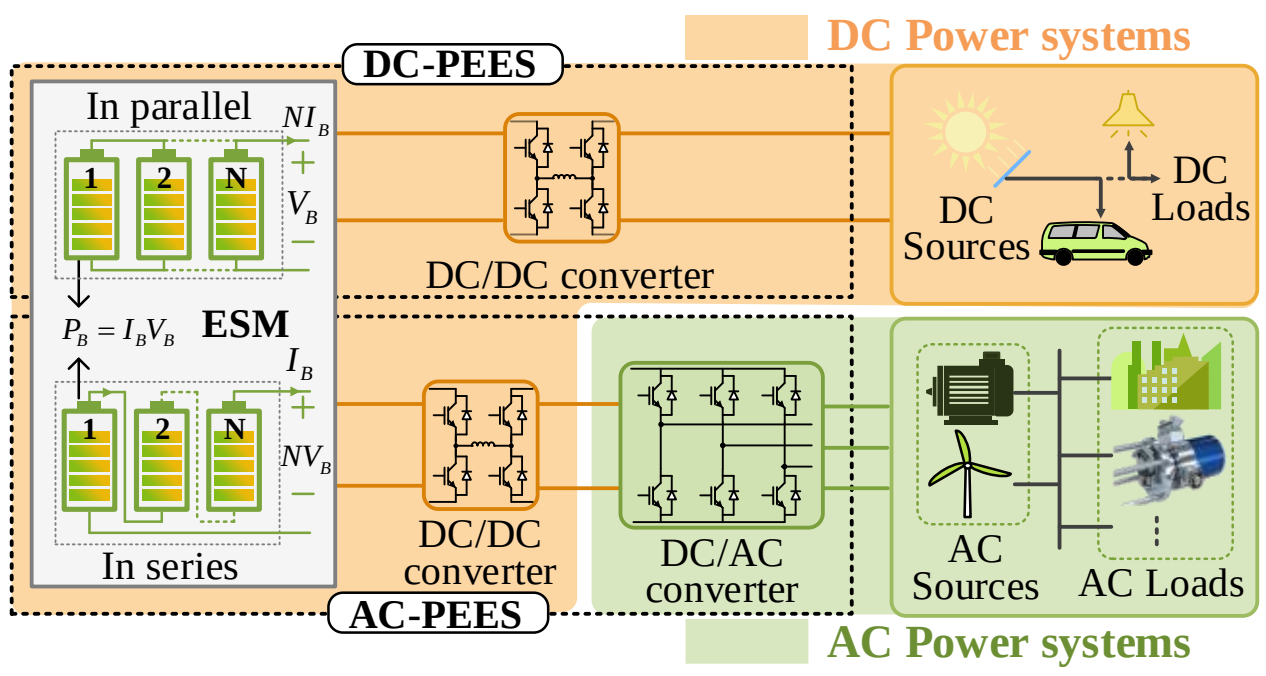

Fig 3. A PEES connected to DC and AC power systems (DC-PEES and AC-PEES).

As depicted in Fig. 3, the PEES is composed of an ESM (as indicated by the gray block), converters, and the connection lines between them. The PEES is classified into DC-PEES and AC-PEES based on whether the PEES is connected to DC or AC power systems. In the DC-PEES, the ESM connects to DC power systems through a DC/DC converter (as shown at the top of Fig. 3), and in the AC-PEES, the ESM connects to AC power systems through a DC/AC converter, with the option of a DC/DC converter (as shown at the bottom of Fig. 3).

In the ESM, numerous submodules are interconnected in series and/or parallel, which can be any type of batteries or other technologies for absorbing/outputting power [56]. To clarify, denote the rated voltage, current, and power of a submodule as $V_B$, $I_B$, and $P_B$, respectively, $N$ submodules can be interconnected either in series or parallel (as presented by batteries in Fig. 3).

In converters, various power electronic topologies can be adopted, for example, mostly adopted topologies of DC/AC converters are two-level [57], three-level [58], and modular multilevel topologies [59]. More details can be found in [60]. However, when focusing on the dynamics of entire power system rather than inner switches, the average value model is typically used [61]. Therefore, redundant introductions to power electronic topologies will not be provided in the following sections, and the average value model serves as the basis for analyzing the dynamics of power systems integrated with PEESs.

### 2.2. Fundamental Control Strategies of DC/DC Converters

Four types of control strategies for DC/DC converters are illustrated in Fig. 4, frequently used when connecting the ESM to DC power systems. In this context, $V_{dc}$, $I_{dc}$, and $P_{dc}$ refer to the DC voltage, current, and active power output from the DC/DC converter. The superscript "*ref*" denotes the reference for control loops. If the values of $I_{dc}$ and $P_{dc}$ are positive, indicating that the power flow aligns with the arrow, the PEES is discharging. Conversely, the PEES charges from the DC power system.

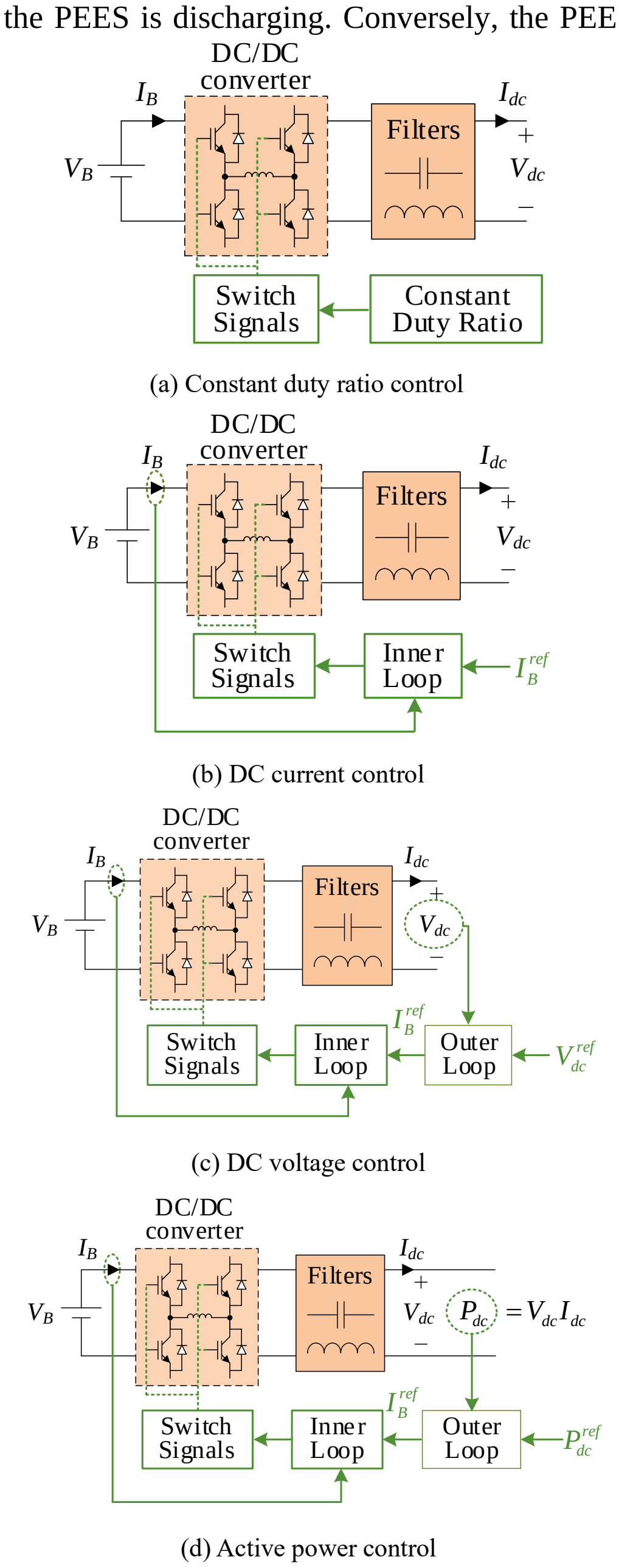

(a) Constant duty ratio control

(b) DC current control

(c) DC voltage control

(d) Active power control

Fig 4. Fundamental control strategies for DC/DC converters

In Fig. 4 (a), the constant duty ratio control can directly set a constant duty ratio between $V_B$ and $V_{dc}$, adjusting the duty ratio can change the outputted DC voltage of the DC/DC converter [62]. To improve the control flexibility, in Fig. 4 (b), DC current control introduces a control loop to the duty ratio. The duty varies as the DC current changes until the detected DC current aligns with its reference value [63]. To further enhance control capabilities, in Fig. 4 (c) and (d), DC voltage and active power controls add an outer control loop to the current control loop. These controls generate the reference of DC current based on the instructions of DC voltage and active power, thereby achieving constant power and DC voltage output [64–66].

These fundamental control strategies change PEESs' steady-state characteristics as desired, which can be different from that of ESM. However, parts of the strategies may potentially worsen dynamic characteristics under specific conditions. Notably, the active power control can provide negative damping to the DC power system, thus leading to growing oscillations [67]. The authors in [68] compared the dynamic characteristics and their impact of different controls on DC power system stability, highlighting that active power control poses a higher instability risk to DC voltage. Therefore, a meticulous design of control loops is crucial to ensure the stable operation of PEESs.

### 2.3. Fundamental Control Strategies of DC/AC Converters

The normally used coordinates of AC power systems include three-phase, $\alpha$-$\beta$, $d$-$q$, $x$-$y$, and sequence-domain coordinates. The corresponding relationship is shown in Fig. 5.

The control strategies of DC/AC converters can be classified into grid-following (GFL) control and grid-forming (GFM) control based on different synchronous mechanisms. The detailed control loops are illustrated in Fig. 6, $V_{xy}$ and $I_{xy}$ denote the AC voltage and current at the Point of Common Coupling (PCC) of the AC power system in $x$-$y$ coordinates (see blue arrows in Fig. 5 for clarification). $V_q$, $V_d$, and $I_q$, $I_d$ are $d$-$q$ components (see green arrows in Fig. 5 for clarification) of AC voltage and current, respectively.

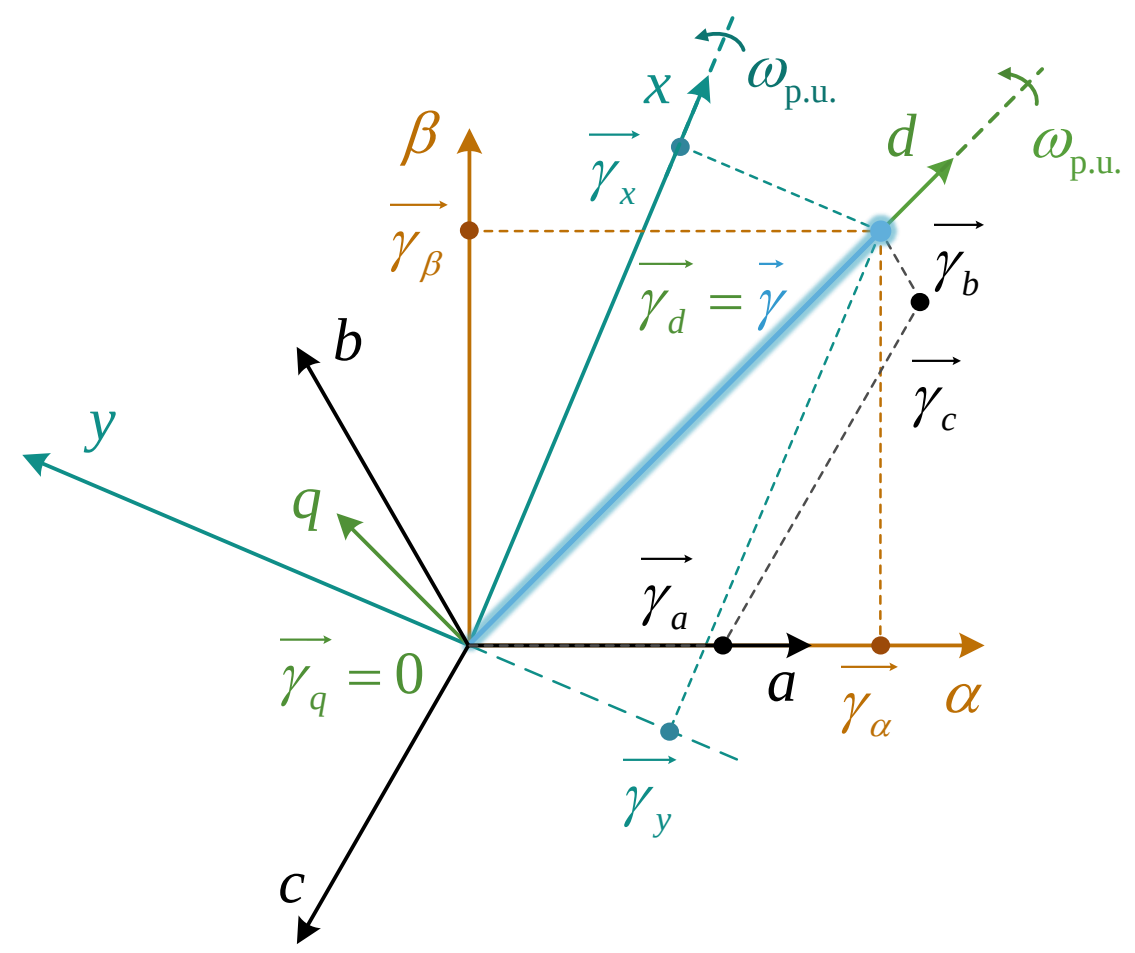

Fig 5. Relationships among three-phase, $\alpha$-$\beta$, $d$-$q$, and $x$-$y$ coordinates.

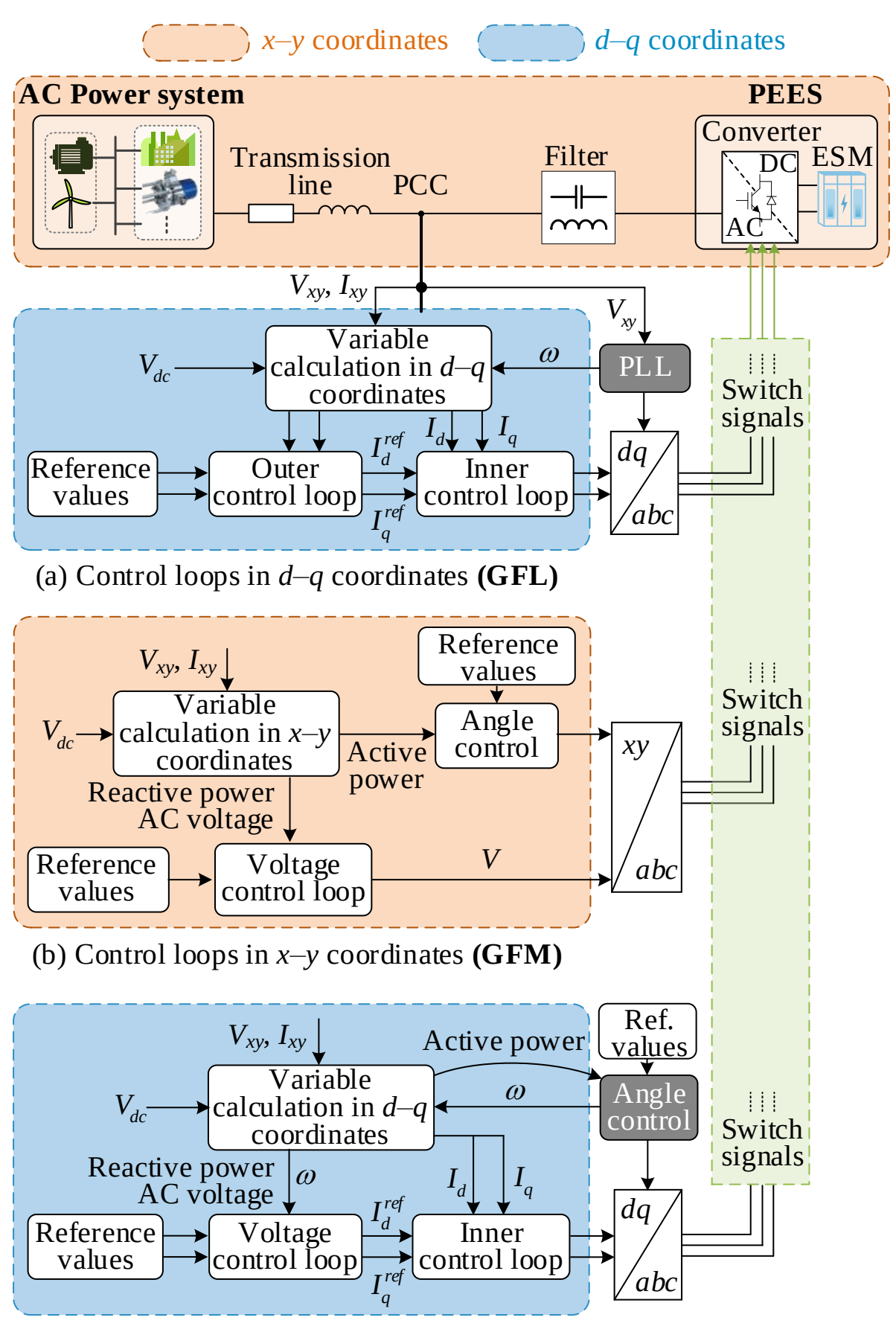

Fig 6. Fundamental control strategies for DC/AC converters.

In Fig. 6 (a), the GFL control comprises three control loops: outer control loop, inner control loop, and phase-locked loop (PLL). The outer control loop is primarily responsible for generating reference values for currents in $d$-$q$ components based on specified references. For example, the active and reactive powers were used as references to maintain a constant power output from the converter [69,70]. In [71] and [72], the frequency and AC voltage were used as references, allowing the converter to participate in the dynamic process of AC power system. To address additional objectives, droop control was used in [73], enabling coordination among multiple variables.

Subsequently, the inner control loop adjusts the output currents to match the current references, concurrently achieving the decoupling of active and reactive power. It is important to note that all these control loops operate in the *d*-*q* coordinates of the DC/AC converter, which are established by PLL, rather than AC generators. Therefore, the PLL serves as a crucial loop for synchronization with the connected power system, and its dynamics require significant attention. For instance, weak connections between the DC/AC converter and the AC power system [74,75] pose potential risks for the stability.

The GFM control typically comprises the voltage amplitude control loop, the voltage angle control loop, and an optional inner control loop. However, the detailed construction of these loops varies depending on the coordinates used. Therefore, we classify GFM control into two types: GFM control established in *x*-*y* coordinates and in *d*-*q* coordinates.

In Fig. 6 (b), these control loops are established in *x*-*y* coordinates, the voltage amplitude and angle control loops output the references of voltage amplitude and angle to the converter, respectively. Thus, the inner control loop is not used, typical configurations were summarized in [76]. By simulating synchronous generators (SGs), the converter can further contribute inertia and damping to the AC power system [77].

In Fig. 6 (c), these control loops are established in *d*-*q* coordinates, the voltage angle control loops output voltage angle (or angular speed), based on which, *d*- and *q*-axes voltages are calculated. The voltage amplitude control loop outputs the references of *d*- and *q*-axes currents to inner control loop. This is achieved by controlling the *d*-axis voltage to the voltage amplitude and the *q*-axis voltage to be zero. It should be noted that in this configuration, a transfer between the *d*-*q* and *x*-*y* coordinates is required, which is a significant difference from that established in *x*-*y* coordinates. The authors in [78] used this configuration to realize the regulation of the frequency and voltage in an isolated power system.

However, regardless of the adopted configuration, the GFM control avoids the use of PLL, thereby eliminating the instability risks associated with PLL. For instance, in [70], the authors compared the impact of GFM control and GFL control on the stability of power systems under weak connections, demonstrating that GFM control can improve power system dynamics. Additionally, as indicated by [79], a GFL controlled converter cannot successfully operate if power systems do not include GFM controlled equipment and synchronous generators.

### *2.4. Differences between PEESs and REs in Construction and Control*

**Table 2**
Differences between PEESs and REs in construction and control

| | Energy generation | Control strategies | Power flow |
|---|---|---|---|
| PEESs | Battery, chemical processes | Various and multi-functional | Bidirectional |
| REs | Wind generators and photovoltaic panels, physical processes | Maximum power output and is limited by environment | Directional |

Although there are many similarities between PEESs and REs, Table 2 highlights three important differences, which suggests that the impact of them on power system dynamics should not be considered identical.

1) Energy generation process: In PEESs and REs, energy is generated in different methods. Although different methods may yield similar results in steady-state, their dynamics differ significantly considering distinct energy generation processes. Therefore, analogous to modeling wind generators [80] and photovoltaic panels [81], establishing dynamic models for ESMs is necessary.

2) Control strategies of DC/AC converters: The active support provided by REs is limited by environmental factors [82], particularly in the case of wind generators and photovoltaic panels. In contrast, PEESs offer better controllability and can actively support the AC power system in multiple aspects, including inertia [83,84], AC voltage [85], and damping [86]. Consequently, dynamics of these control strategies should be carefully reconsidered to avoid potential risks at early stage.

3) Power flow direction: Traditional instability risks have primarily focused on directional power flow of REs, however, the reversed power flow of PEESs may induce new risks. Therefore, it becomes crucial to address the potential risks associated with power flow reversals in large-scale PEESs interconnected power systems.

Considering these differences, it is essential to develop specific models for PEESs with various control strategies, analyze the impact of unique dynamics and bidirectional power flow capabilities. This will contribute to a more accurate understanding of the dynamic characteristics of PEESs and ensure stable integration into power systems.

## 3. Dynamic Modelling of PEESs

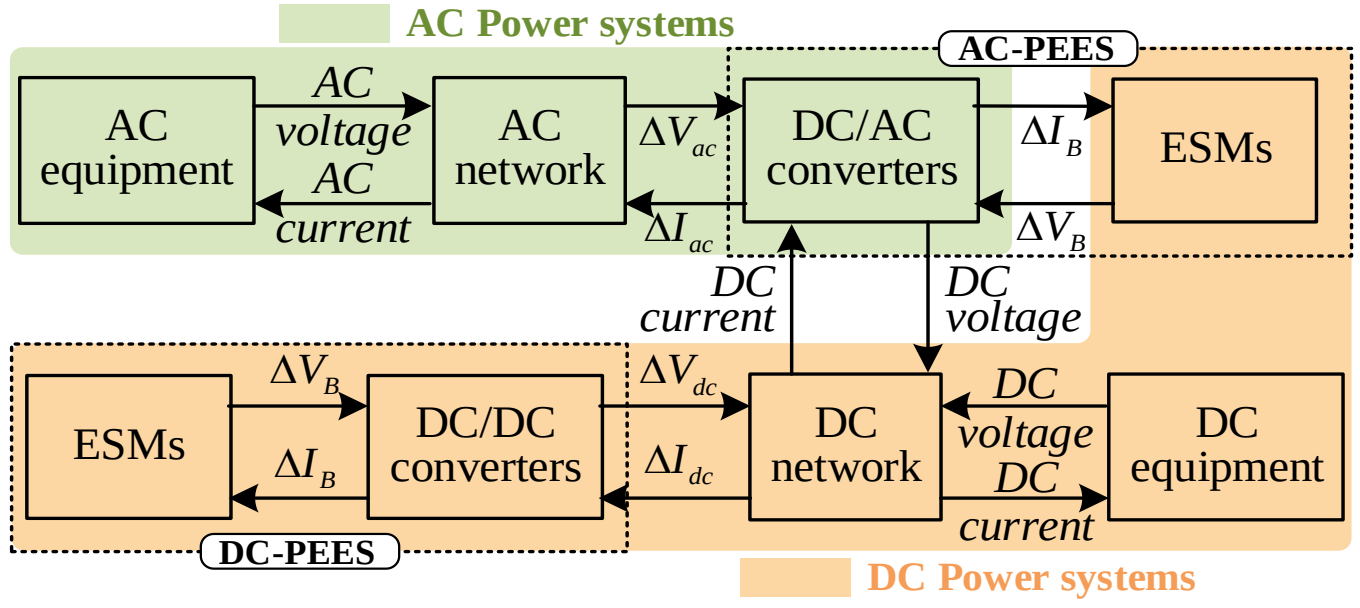

Fig 7. Illustration of modular modeling method for AC and DC power systems connected with PEESs.

The modular modeling method has been a primary choice for establishing dynamic models in complex power systems [87]. With that, the model of PEESs can be constructed by combining models of a series of modules, as illustrated by Fig. 7. It addresses the challenge of high dimensionality when establishing models for complex power systems in a direct way, making the modeling process more manageable. Therefore, in this study, the model of PEESs is divided into three modules, that is, ESMs, DC/DC converters, and DC/AC converters.

### *3.1. Modeling of ESMs*

Based on dynamic processes of ESMs, the mathematical equations between $V_B$ and $I_B$ can be obtained. The dynamic model of ESMs is then derived by linearizing them as:

$$\Delta V_B = G_{ESM}(s)\Delta I_B \tag{1}$$

where, $G_{ESM}(s)$ is a transfer function that represents the dynamics of ESMs.

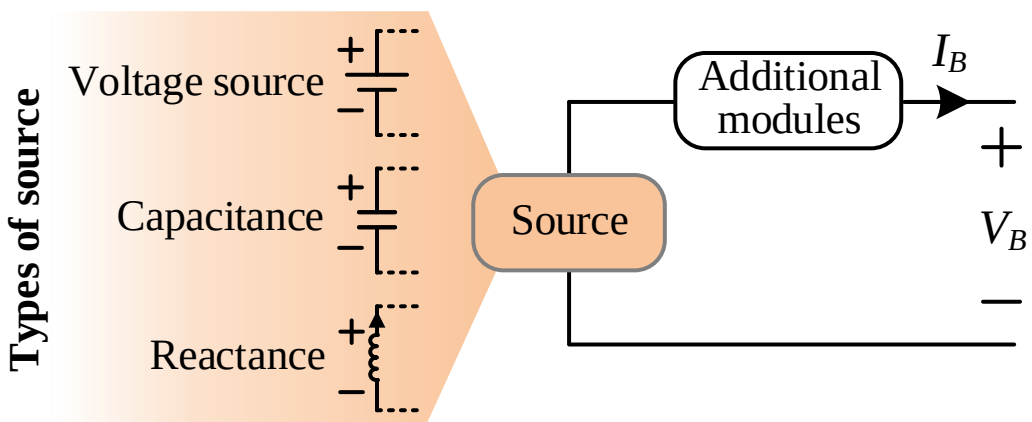

Fig 8. Common modeling diagrams of ESMs.

**Table 3**
Summary of detailed models of ESMs

| Types | Sources | Additional modules | Refs |
| --- | --- | --- | --- |
| Supercapacitor | Capacitance | Resistance | [88] |
| Superconducting magnetic module | Reactance | \ | [89] |
| Fuel cell | Voltage source | Resistance and capacitance | [90] |
| Lithium-ion battery | Voltage source | Resistance and capacitance | [91] |

The modeling of supercapacitor, fuel cells, superconducting magnetic energy storage, and thermal cells should be distinct to match their dynamic characteristics. For instance, in [88], the supercapacitor model was established, dynamics were represented by variated capacitances and a series resistance. In [89], dynamics of superconducting magnetic module were represented by a reactance. In these cases, the voltage varies instead of being constant when power varies. In [90], the fuel cell model was introduced, with resistance and capacitance added to a voltage source, differing from [89] and [92]. In [91], a lithium-ion battery model was introduced, adding resistance and capacitance to the voltage source, which differs only in details from [90]. Fig. 8 shows common modeling diagrams of ESMs, and details are summarized in Table 3.

The electrochemical battery is the most commonly used in various types of ESMs [93]. Developing an accurate model for the battery is a mainstream task. This is challenging as it involves various factors such as material characteristics, operational status, and external conditions. Several studies have considered detailed factors to obtain more accurate voltage-current characteristics of batteries, such as the work in [94] considered the dependence of battery charging ability on its state of energy, [95] considered battery current error to accurately estimate SOC, and [96] considered the deviation characteristics, and provided detailed modeling. These studies contribute to understanding the detailed dynamics of ESMs, promoting their full-order modeling.

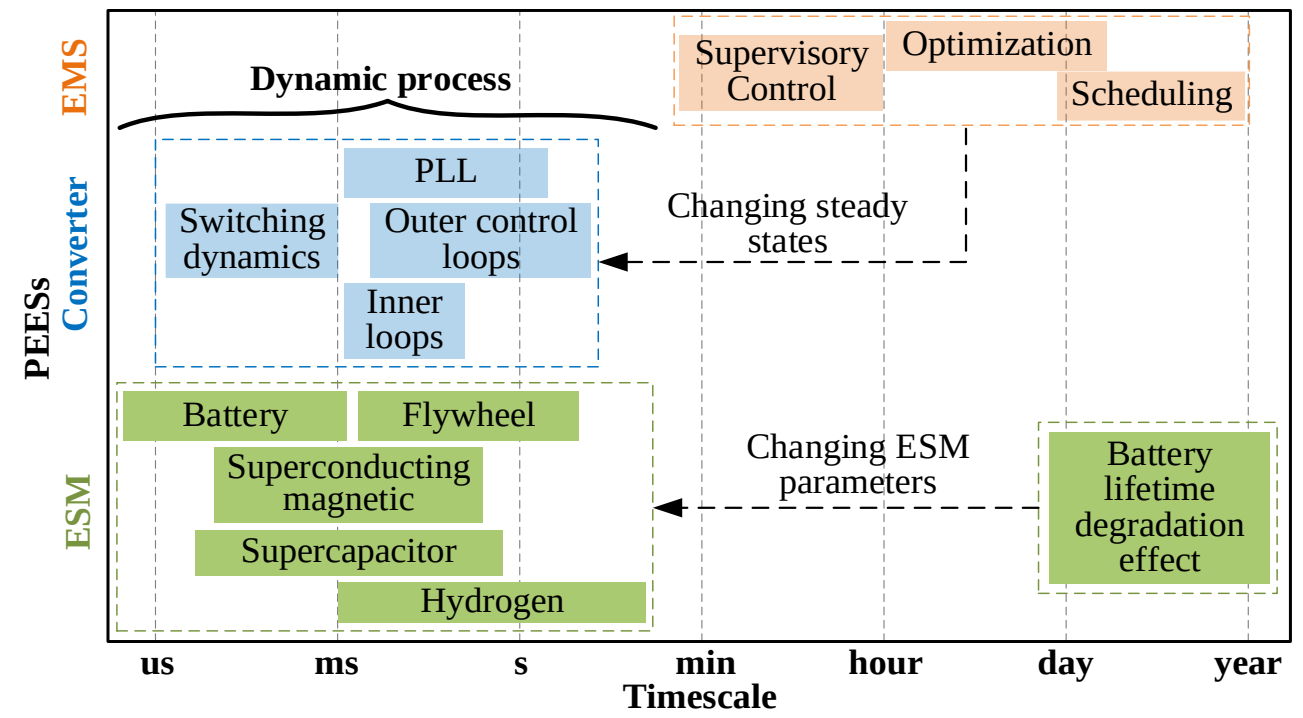

Fig 9. Timescale of different parts of PEESs.

As illustrated in Fig. 9, the dynamics of PEESs are affected by three main components: ESM, converter, and Energy Management Systems (EMS). These components operate on different timescales, resulting in varying response times [97]. The EMS is responsible for optimizing operational states at the highest level [98], such as scheduling [99,100] optimization [101–103], and supervisory control [104]. As the instructions are determined and outputted to the converter, the control loops act to track these instructions. At a given moment, the dynamic response of PEESs is unrelated to the EMS, as the EMS operates on a longer timescale. However, changes in the EMS can alter the steady state of the PEESs, which in turn may affect the dynamic response at that moment.

Different control loops have different dynamic response speeds, corresponding to different timescales as indicated by blue blocks in Fig. 9. However, the achievement of instructions also depends on the ESMs. Among them, batteries typically have a faster power response, enabling them to respond to control loop instructions more quickly. On the other hand, flywheels have slower power response times.

However, many dynamics are often neglected to simplify the dynamic model of ESMs, a commonly used simplified model for dynamic analysis is a voltage source in series with a resistance [105]. Based on this simplified model, the impact of ESs on black start was studied in [106], the impact of DC lines and control parameters was analyzed in [107] and [108], respectively. It should be noted that the simplified model is not universally effective under all conditions. Errors may be amplified, affecting the accuracy of analyzed results if the simplified model does not perfectly capture the dynamic characteristics of ESMs. For example, in [109], the authors demonstrated that battery degradation increases the cost of battery life loss, while research [110] described the capacity loss due to cycling. Both of the studies indicate that the degradation of battery lifetime on a larger timescale can alter the dynamic characteristics of batteries by changing the inner parameters of the battery. Therefore, when using the simplified model, the conditions should be carefully evaluated to determine whether the parameters are affected by long-term factors, in order to obtain precise results. This consideration has not always been rigorously addressed in previous studies.

### *3.2. Modeling of DC/DC Converters*

The dynamic characteristics of DC/DC converters mainly depend on their control loops, as represented by:

$$\Delta V_{dc} = G_{DC1}(s)\Delta I_{dc} + G_{DC2}(s)\Delta V_B \tag{2}$$

In (2), ESMs affect the dynamics of DC/DC converters by changing the response of $\Delta V_B$, and $G_{DCi}(s)$, $i$ = 1, 2, represents the dynamics of DC/DC converters, mainly related to their control loops. Specially, if the ESM is simplified as a constant voltage source ($\Delta V_B = 0$), $G_{DC1}(s)$ is an input impedance of the DC/DC converter.

It is important to note that any modification to control loop necessitates updating the details of $G_{DCi}(s)$. For example, in the case of constant duty ratio control, its model was derived in [111]. Regarding commonly used control loops, the model of a DC/DC converter adopting constant DC voltage control was derived in [112], while the model for a DC/DC converter adopting active power control was established in [113]. Additionally, the dynamics of special controls of DC/DC converters may require independent analyses. For instance, in [114], the dynamics of a bilateral DC/DC converter were analyzed using a time-domain model.

To reduce the complexity in dynamic analysis, some studies simplify the model under specific conditions. In [115], the DC/DC converter using DC voltage control was simplified as a constant DC voltage. In [68,116], the DC/DC converter using current control was simplified as a current source. In [115], the DC/DC converter using active power control was simplified as a constant power. These simplifications are made under the condition that the dynamics of control loops are much faster or slower than the aimed dynamics. Otherwise, the simplifications may cause significant errors, making the results unreliable. Therefore, the model verifications and error analyses should not be skipped when using simplified models.

### *3.3. Modeling of DC/AC Converters*

The representation of dynamic characteristics of DC/AC converters is similar to that of DC/DC converters, also depend on control loops and ESMs, yields:

$$\Delta I_{ac} = G_{AC1}(s)\Delta V_{ac} + G_{AC2}(s)\Delta V_B \tag{3}$$

In (3), $\Delta I_{ac}$ and $\Delta V_{ac}$ are vectors, and $G_{AC1}(s)$ is a multi-dimension transfer function, which depends on the adopted coordinates. As illustrated in Fig. 5, the normally used coordinates of AC power systems include three-phase, *α-β*, *d-q*, *x-y*, and sequence-domain coordinates. In three-phase coordinates, $\Delta I_{ac}$ and $\Delta V_{ac}$ include variables in phases a, b, and c. Consequently, $G_{AC1}(s)$ is of three orders and exhibits cross-coupling among phases. The detailed model can be referred to in [117]. Under the assumption of three-phase symmetry, the vectors of $\Delta I_{ac}$ and $\Delta V_{ac}$ can be reduced to two orders by modeling in *α-β*, *d-q*, *x-y*, and sequence-domain coordinates.

The transformation from three-phase to *α-β* coordinates is given by

$$\begin{bmatrix} V_\alpha \\ V_\beta \end{bmatrix} = \frac{2}{3}\begin{bmatrix} 1 & -\frac{1}{2} & -\frac{1}{2} \\ 0 & \frac{\sqrt{3}}{2} & -\frac{\sqrt{3}}{2} \end{bmatrix}\begin{bmatrix} V_a \\ V_b \\ V_c \end{bmatrix} \tag{4}$$

In (4), $V_a$, $V_b$, and $V_c$ represent time-variable voltage components on stationary axes *a*, *b*, and *c* in three-phase coordinates.

$V_\alpha$ and $V_\beta$ denote time-variable components on stationary axes $\alpha$ and $\beta$ in $\alpha$-$\beta$ coordinates. In [118], a mathematical model of battery charging in $\alpha$-$\beta$ coordinates was developed and presented. Additionally, in [119], a unified model predictive control scheme for an integrated photovoltaic and battery storage system was proposed based on the model in $\alpha$-$\beta$ coordinates. However, analyzing variables in $\alpha$-$\beta$ coordinates is still challenging due to their time variability. To address this issue, models in *d-q* and *x-y* coordinates (as indicated by green and blue arrows in Fig. 5.) receive more attention.

The transformation from three-phase to *d-q* and *x-y* coordinates is given by

$$\begin{bmatrix} V_x \\ V_y \end{bmatrix} or \begin{bmatrix} V_d \\ V_q \end{bmatrix} = \begin{bmatrix} \cos(\omega t) & \sin(\omega t) \\ -\sin(\omega t) & \cos(\omega t) \end{bmatrix} \begin{bmatrix} V_\alpha \\ V_\beta \end{bmatrix} \tag{5}$$

In (5), $V_d$ and $V_q$ denote constant components on rotating axes *d* and *q* in *d-q* coordinates of DC/AC converters, while $V_x$ and $V_y$ denote constant components on rotating axes *d* and *q* in *d-q* coordinates of the rotator of synchronous generators. To distinguish between them, the latter is named *x-y* coordinates. $\omega$ represents the angular speed output by the PLL for *d-q* coordinates, whereas it is the same as the angular speed of the rotator of synchronous generators for *x-y* coordinates. At a certain node of the power system, these components can be transferred to each other [120]. For the dynamic analysis of grid-connected DC/AC converters, a state-space model was derived in [121], and an impedance model was derived in [112], both in *d-q* coordinates of DC/AC converters. To connect DC/AC converters to AC network, the transfer between *d-q* and *x-y* coordinates was considered in [122], resulting in the establishment of the model in *x-y* coordinates. However, in some cases, there is coupling between the *d* and *q* axes. To decouple them and simplify analysis, models in sequence-domain coordinates are derived.

The transformation from *d-q* (*x-y*) to sequence-domain coordinates is given by

$$\begin{bmatrix} V_p \\ V_n \end{bmatrix} = \frac{1}{\sqrt{6}} \begin{bmatrix} V_d + jV_q \\ V_d - jV_q \end{bmatrix} \tag{6}$$

In (6), $V_p$ and $V_n$ denote constant voltage components on positive and negative sequences, respectively. For instance, in [123], a model of the DC/AC converter in the sequence-domain was established. It's important to note that models in different coordinates can be transformed into each other [123,124] as illustrated in Fig. 5. Therefore, the dynamics represented by these models are equivalent, and different model choices cannot change the results. However, considering specific scenarios, proper model selection can simplify the dynamic analysis process.

Similarly, simplified models of DC/AC converters are widely employed to mitigate the complexity arising from high-order dynamics. Studies in [125] indicated that the dynamics of DC/AC converters using active and reactive power control can be simplified as a current source with high impedance in parallel, while those using AC voltage and frequency control can be represented by a voltage source with a small impedance in series. In [126], slower dynamics of outer control loops were ignored, focusing on the fast dynamics of filters. Likewise, in [127], fast dynamics of inner current control loops were neglected as the focus is on the subsynchronous frequency range. Obviously, there are limitations to the application of simplified models that need to be considered carefully. The details of module models are summarized in Table 4.

**Table 4**
Summary of module models

| Modules | Types | Detailed model | Simplified model |
|---|---|---|---|
| ESMs | Supercapacitor | [92] | [105–108] |
| | Superconducting magnetic ES | [89] | |
| | Fuel cell | [90] | |
| | Lithium-ion battery | [91,94–96] | |
| DC/DC converter | Duty ratio control | [111] | [68,108,115] |
| | DC voltage control | [112] | |
| | DC current control | [63] | |
| | Active power control | [113] | |
| DC/AC converter | Three-phase coordinate | [117] | [125–127] |
| | *α-β* coordinate | [118,119,124] | |
| | *x-y* coordinate | [120,122,124] | |
| | *d-q* coordinate | [112,120,121,124] | |
| | Sequence-domain | [123,124] | |

### 3.4. *Error Evaluation of Simplified Model*

The development of a simplified model results from finding a balance between accurate dynamic analysis of power systems integrated with large-scale PEESs and the limited computational resources. In different scenarios, the required level of detail of a model varies, for example, in [128], when analyzing the impact of dynamic interactions between converters on stability, the dynamics of the AC system and inner-loop control are neglected to reduce the model order, focusing instead on the dynamics of the outer-loop controller and the current flow controller (CFC); in [54], when investigating the effects of power supply switching ripple and losses on the state of ESs and their charging and discharging characteristics, it is essential to take into account the internal parameters of the battery models, or the results may be inaccurate; in [52], during the stability research conducted under normal operating conditions, simplifying DC/DC or DC/AC converters does not significantly affect the stability analysis results but only impacts the Total Harmonic Distortion (THD), making it a viable approach to reduce the complexity of the analysis. When the number of PEESs is relatively low, the errors introduced by simplified models are negligible in terms of their impact on the results. However, as the scale of PEESs expands, errors caused by individual PEESs may aggregate, leading to "error aggregation". This aggregation may result in inaccuracies in the analyzed results of large-scale PEESs, where the authors highlight the misconception of assuming control dynamics to be ideal [52]. Therefore, it is essential to pay more attention to propose a numerical evaluation method for errors. This approach ensures that errors are additionally considered in analyzed results, thus improving the reliability of the analyzed conclusions. In the following sections, explanations of error aggregation and the idea for proposing error numerical evaluation method are introduced.

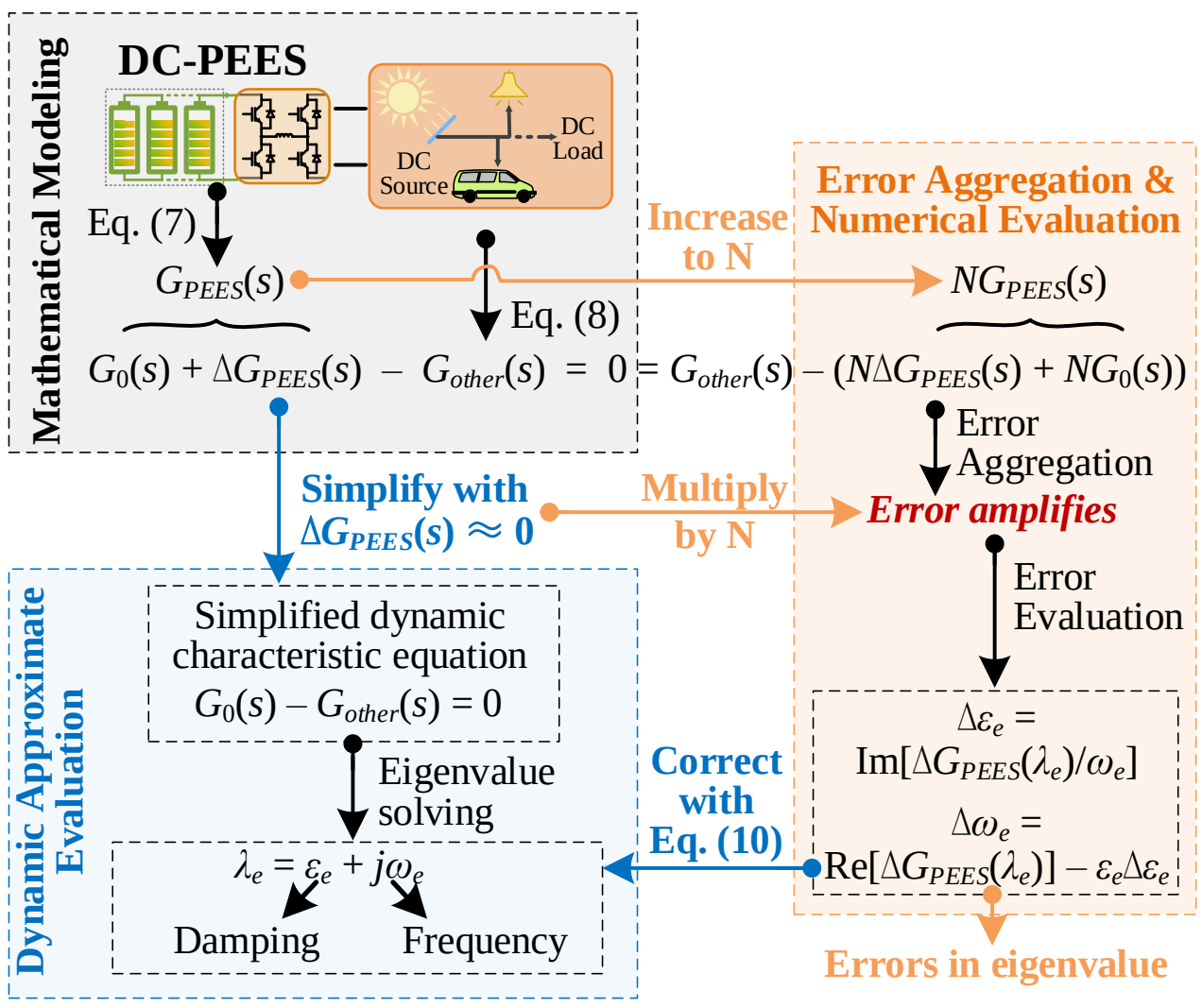

Fig 10. Illustration of numerical evaluation of model errors.

Denote the model of a PEES connected to the DC power system as $G_{PEES}(s)$ from (1) and (2), that is,

$$\Delta V_{dc} = G_{PEES}(s)\Delta I_{dc} \tag{7}$$

where, the active power injected into and output from the DC/DC converter is the same, such that $\Delta(V_{dc}I_{dc}) = \Delta(V_B I_B)$,

$$G_{PEES}(s) = \frac{G_{DC1}(s) + \dfrac{V_{dc}G_{DC2}(s)G_{ESM}(s)}{V_B + I_B G_{ESM}(s)}}{1 - \dfrac{I_{dc}G_{DC2}(s)G_{ESM}(s)}{V_B + I_B G_{ESM}(s)}}.$$

The model of the remaining DC power system is defined as $G_{other}(s)$, satisfying

$$\Delta V_{dc} = G_{other}(s)\Delta I_{dc}. \tag{8}$$

From (7) and (8), the dynamic characteristic equation of the DC power system connected by a PEES is obtained as (9), and illustrated in the gray block in Fig.10:

$$G_{PEES}(s) = G_0(s) + \Delta G_{PEES}(s) = G_{other}(s). \tag{9}$$

In (9), $G_0(s)$ and $\Delta G_{PEES}(s)$ represent the simplified model and the error between the accurate and simplified models. The dynamics of the DC power system connected with PEES can be analyzed by the eigenvalues, i.e., solutions, of (9). Stability holds only when all solutions are located on the left side of complex plane. Generally, $\Delta G_{PEES}(s)$ is small because the simplified model is approximately close to the accurate model of a PEES. The approximation evaluation of the dynamics of the power system can be conducted as depicted in the blue block in Fig. 10.

However, if PEES expends, the aggregation of $\Delta G_{PEES}(s)$ may be large, potentially leading to miscalculations of the solutions, as illustrated in the orange block in Fig. 10. To address this issue, the accuracy of calculated results can be improved by considering the impact of the errors caused by simplified models, where the modal analysis is presented as an example.

The errors in damping ($\Delta\varepsilon_e$) and frequency ($\Delta\omega_e$) of the oscillation is evaluated by using the damping torque analysis method [128] as

$$\begin{aligned} \Delta\varepsilon_e &= \frac{\text{Im}[\Delta G_{PEES}(\lambda_e)]}{\omega_e} \\ \Delta\omega_e &= \text{Re}[\Delta G_{PEES}(\lambda_e)] - \varepsilon_e \frac{\text{Im}[\Delta G_{PEES}(\lambda_e)]}{\omega_e} \end{aligned} \tag{10}$$

In (10), $\lambda_e = \varepsilon_e + j\omega_e$ is the dominant oscillation mode when simplified model is used. The accuracy of analyzed results improves when the impact of errors calculated in (10) are added. However, this is a tutorial case of evaluating the model errors, and there are many potential methods that can achieve this aim, requiring further research to explore. With this method, the calculation does not significantly increase, and the accuracy improves, which is especially suitable for the dynamic analysis of large-scale PEESs integrated power system in the future.

## 4. Dynamic Instability Risk Assessments of DC Power Systems Integrated with PEESs

### *4.1. Dynamic Instability Risk Assessments Considering a Single DC-PEES*

#### *4.1.1. Results of Dynamic Instability Risks*

The instabilities of PEESs are from two aspects. The first aspect is the self of PEESs when they are not connected to DC power systems. In [112], the self-stability of PEES was analyzed using the Routh stability criterion, and proper values for control parameters were determined. However, the self-stability of PEESs has been ensured in normal because the power system cannot allow the integration of an unstable equipment. Therefore, the major instability risks of PEESs arise from the second aspect, that is, dynamic interactions between the PEES and the connected DC power systems [128].

In [129], the charging and discharging dynamics of ESMs were considered, and the impact of ESM parameters on stability was analyzed using the modal analysis method. In [130], the internal resistance of ESM was considered, indicating that excessive charging and discharging of ESM can increase internal resistance and decrease battery voltage, thereby worsening stability. To simplify the analysis, many studies ignore the dynamics of ESMs and represent them as a constant voltage source. The focus is primarily on the impact of DC/DC converter related parameters on stability [50]. In [107], the length of DC lines of the DC/DC converter for grid connection was considered, indicating that stability reduces as the length of DC lines increases. In [68], the impact of different control strategies, current, voltage, and power controls, on stability was compared, indicating that active power control poses higher risks to stability than the others. However, not all control strategies of PEESs may threaten stability, for example, in [131], the control loops were modified by adding feedforward control loops, which improves stability.

#### *4.1.2. Explanations on the Impact of Bidirectional Power Flow*

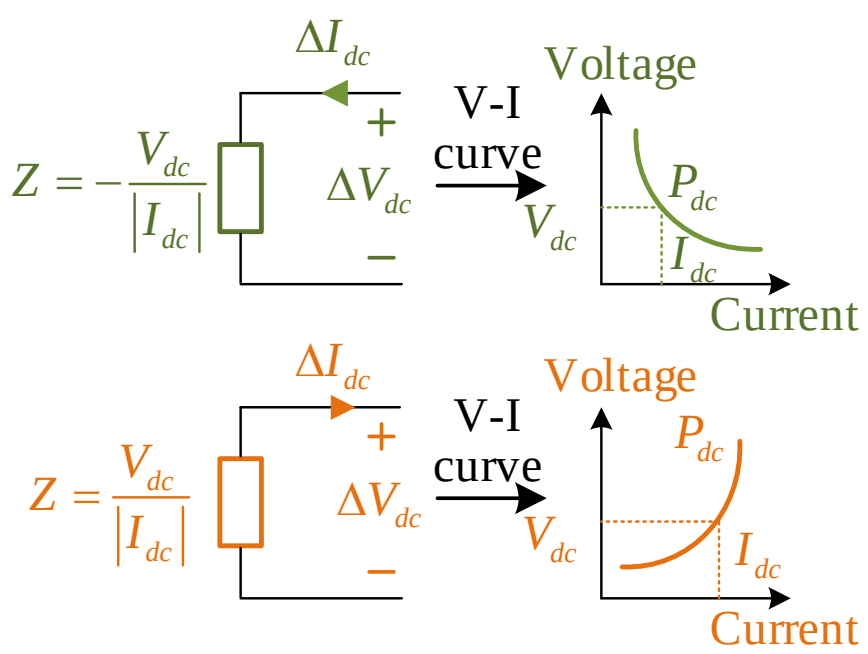

Fig 11. Illustration of the impact of bidirectional power flow of DC-PEESs.

The bidirectional power flow has complex effects on the stability of DC power systems [132], which can be numerically analyzed by modal analysis, frequency-domain analysis, and time-domain simulation methods. However, the analytical equations for these methods are complex, making it difficult to reveal the mechanism of instability. Therefore, in order to provide a clear presentation of how and why the power flow direction can affect the stability when a PEES is connected to DC power systems, PEESs with current control are represented by a current source/load, those with DC voltage control are represented as a DC voltage source, and those with active power control are represented as a constant power source/ load [115].

In theory, using the ideal voltage source cannot cause instabilities since the DC voltage is always constant in this case [68]. However, when the PEESs operate in charging state, constant power control may cause instability due to its negative impedance characteristics [67]. Conversely, if the PEESs operate in the discharging state, stability may improve, as illustrated in Fig. 11. This is a typical case where the power flow of PEES significantly change stability, which holds only under the following condition: the DC-PEES should adopt constant power control, and it can be approximately equivalent to constant power, thus $\Delta P_{dc} = 0$. As there are various control strategies in addition to the fundamental control methods [64], more attention should be paid to the impact of bidirectional power flow caused by PEESs.

### *4.2. Issues and Solutions of Dynamic Analysis for Large-scale DC-PEESs*

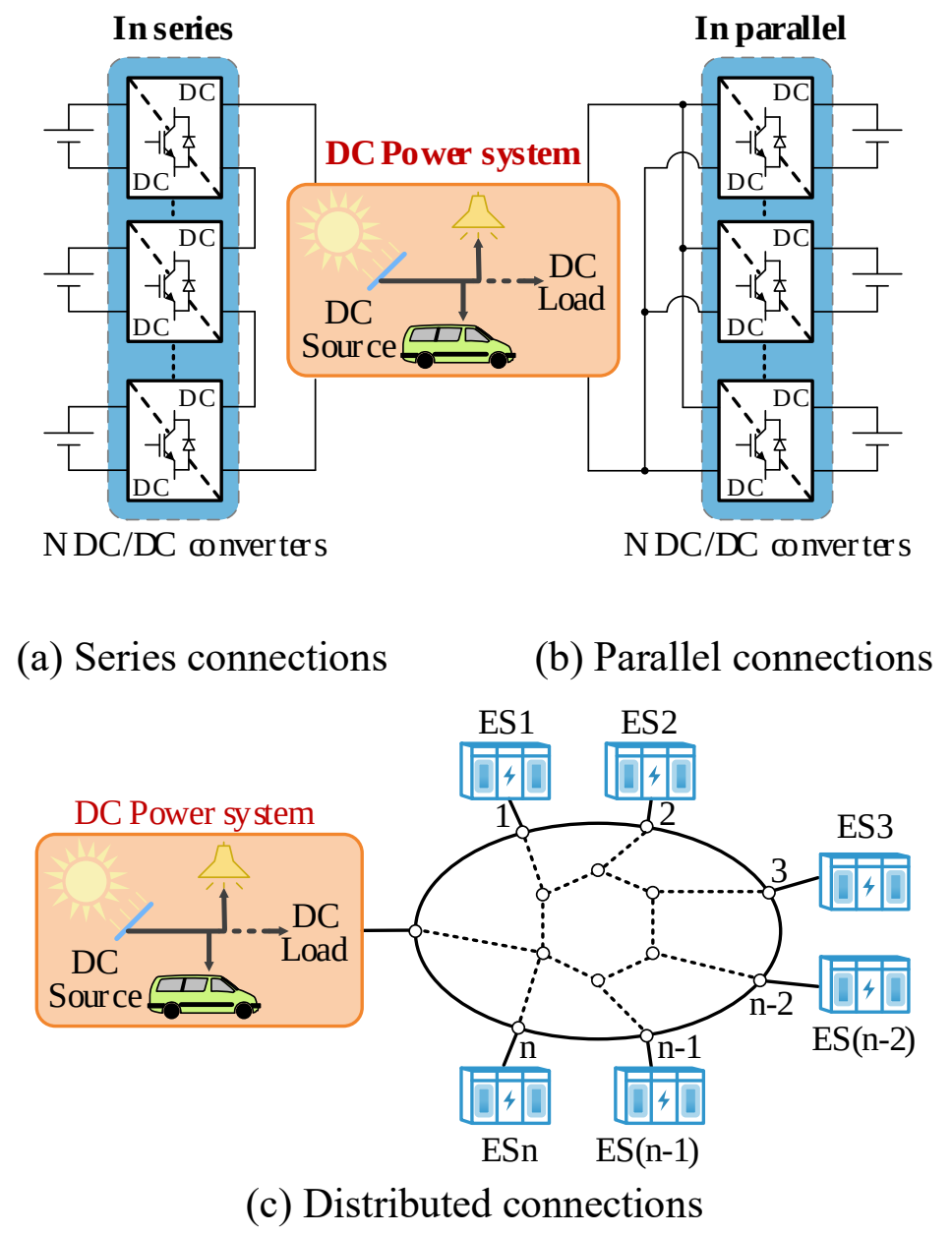

(a) Series connections (b) Parallel connections

(c) Distributed connections

Fig 12. DC power systems integrated with large-scale PEESs.

As shown in Fig. 12, there are three types of connections for PEESs integrating to DC power systems: series, parallel, and distributed. The series connections were illustrated and analyzed in [133,134], clarifying the dynamic relations between the DC voltage and current, and thus verifying the reliability of series connections. The parallel connections were analyzed in [135], where the model is established, and thus the instability risks of this connection are clarified. For these scenarios, the dynamic analysis process is similar to that of a single PEES because the transfer function of PEESs can be linearly obtained with that of a single PEES [112]. Considering that PEESs in series and parallel connections are very similar, the transfer function of $N$ series-connected PEESs can be written as $NG_{PEES}(s)$, and that of $N$ parallel-connected PEESs can be written as $G_{PEES}(s)/N$, with the transfer function of each PEES being $G_{PEES}(s)$.

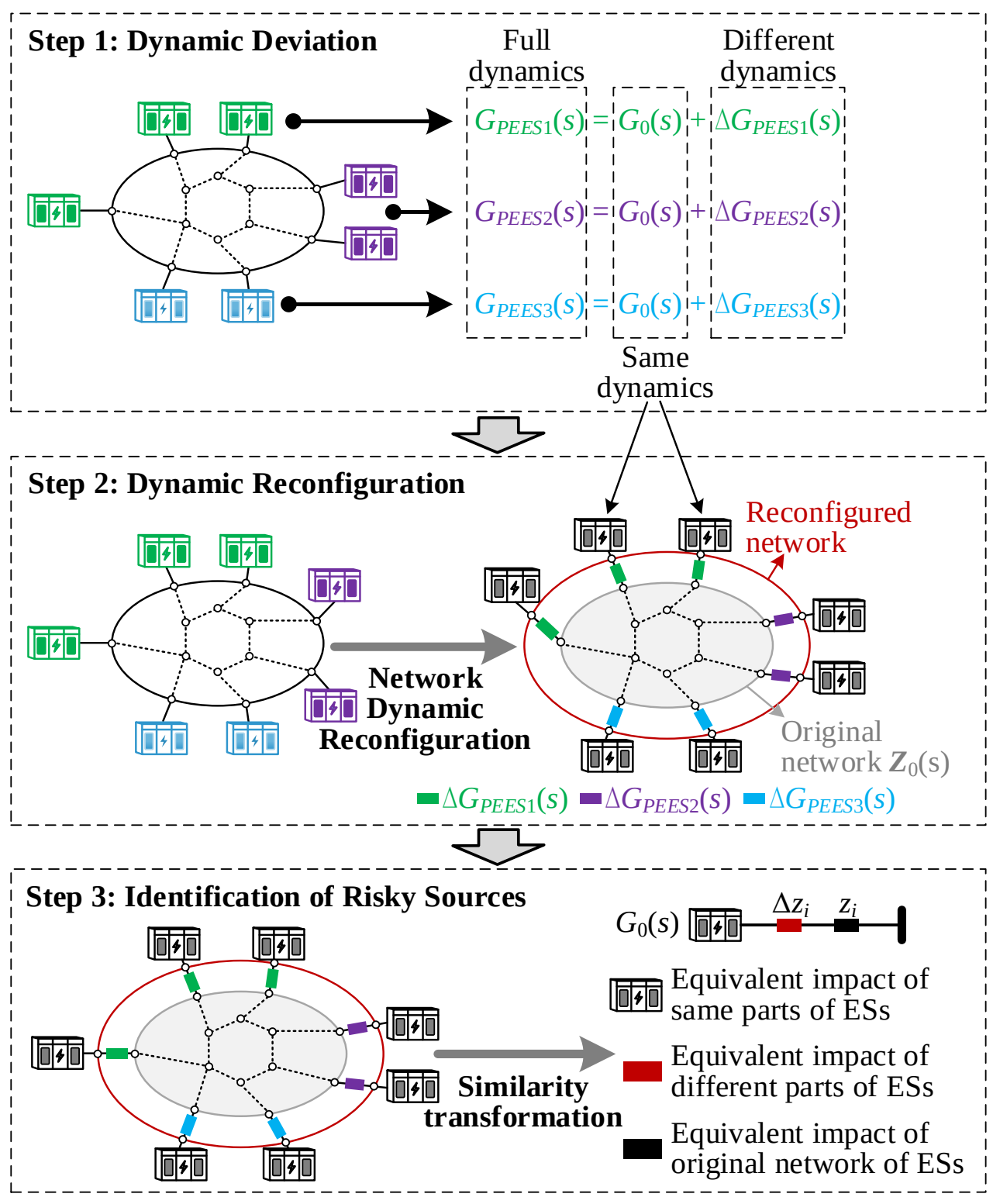

Fig 13. Illustration of process of dynamic reconfiguration.

However, if the PEESs are connected to the DC power system in a distributed way and each PEES is different from another, the equivalent method introduced is invalid. In this case, the interaction variables between PEESs and the DC power system are multiple. Traditional dynamic analysis methods for stability mechanism analysis, e.g., impedance-based and damping-based analysis methods, are inapplicable, as these methods require the interaction variables between the PEESs and the DC power system to be single or equivalent to be single. Although instability risks can still be obtained by modal calculation and time-domain simulation, the reason why this instability occurs is difficult to clarify using the numerical calculation methods. Therefore, the challenge of analyzing the instability of DC-PEESs using distributed connections is the lack of a mechanism analysis method to reveal why this system can be unstable.

To address this issue, this study proposes an idea, namely dynamic reconfiguration [136], which can decompose a complex issue into multiple easier issues, thus determining the stability by investigating these sub-issues. The detailed steps are illustrated in Fig. 13, and explained as follows:

In Fig. 13, three different types of PEESs are considered and represented by different colors. Firstly, the dynamics of different PEESs are divided into identical and different parts, mathematically represented by $G_0(s)$ and $\Delta G_{PEESi}(s)$, $i = 1, 2, 3$, respectively, as given by step 1 in Fig. 13.

Secondly, the different parts (represented by $Diag[\Delta G_{PEESi}(s)]$, $Diag$ represents the matrix as a diagonal matrix) are combined with the original network (represented by $\mathbf{Z}_0(s)$) to form a reconfigured network (as indicated by the red circle in Fig. 13). The dynamic transfer function of the reconfigured network can be obtained as $\mathbf{Z}_0(s)$ - $Diag[\Delta G_{PEESi}(s)]$. Then, the dynamic transfer function model of the DC power system can be obtained as

$$\begin{gathered}\overbrace{Diag\left[G_0(s)\right]+Diag\left[\Delta G_{PEESi}(s)\right]}^{Different\ PEESs}=\overset{Original\ Network}{\mathbf{Z}_0(s)}\\ \Downarrow\\ \overbrace{Diag\left[G_0(s)\right]}^{Identical\ PEESs}=\overbrace{\mathbf{Z}_0(s)-Diag\left[\Delta G_{PEESi}(s)\right]}^{Reconfigured\ Network}\end{gathered} \quad . \tag{11}$$

Regarding the identical parts, dynamics can be analyzed using the similarity transformation method, as verified in [137], where the major instability risks can be identified by equating a power system with multiple PEESs to a power system with a PEES with an equivalent $z_i$, as shown in step 3 in Fig. 13. Regarding the different parts, their impact can be equivalently analyzed with an equivalent additional impedance $\Delta z_i$, where $(z_i + \Delta z_i)$ is the $i^{th}$ eigenvalue of $\mathbf{Z}_0(s)$ - $Diag[\Delta G_{PEESi}(s)]$. The mathematical verification is:

$$\begin{gathered}\mathbf{P}Diag\left[G_0(s)\right]\mathbf{P}^{-1}=\mathbf{P}\left[\mathbf{Z}_0(s)-Diag\left[\Delta G_{PEESi}(s)\right]\right]\mathbf{P}^{-1}\\ \Downarrow\\ Diag\left[G_0(s)\right]=\mathbf{P}\mathbf{Z}_0(s)\mathbf{P}^{-1}-\mathbf{P}Diag\left[\Delta G_{PEESi}(s)\right]\mathbf{P}^{-1}\\ \Downarrow\\ Diag\left[G_0(s)\right]=Diag\left[z_i(s)\right]-Diag\left[\Delta z_i(s)\right]\\ \Downarrow Select\ the\ i^{th}\ subsystem\\ G_0(s)=z_i(s)-\Delta z_i(s)\end{gathered} \quad . \tag{12}$$

where $\boldsymbol{P}$ is an eigenvalue vector matrix.

Finally, the impact of the identical and different parts, as well as the network, on dynamics can be analyzed by selecting one of the subsystems to clarify the major factors of instability based on the simple system. By collecting results from these analyses, the reasons for and mechanisms behind PEESs instability can be elucidated.

## 5. Dynamic Instability Risk Assessments of AC Power Systems Integrated with PEESs

### *5.1. Dynamic Instability Risk Assessments Considering a Single AC-PEES*

#### *5.1.1. Results of Dynamic Instability Risks when AC-PEES Uses GFL Control*

Similar to DC-PEESs, the major factors in AC-PEESs that affect stability come from two parts: the ESMs and DC/AC converter control loops. In some cases, the dynamic characteristics of ESMs can reduce the stability of PEESs. For instance, [138] indicates that inconsistency issues in battery systems can worsen stability. However, more studies simplify the research by considering ESMs as a constant DC voltage [105]. In this latter case, the primary instability risks arise from the control of DC/AC converters, which are summarized in Table 5 and introduced below.

The control of DC/AC converters can be classified into two types: GFL and GFM controls [139]. The instability risks of AC-PEESs are similar to those of REs when the DC/AC converters use GFL-based DC voltage control and output power to the connected power systems. In this case, conclusions drawn from REs can be applicable to AC-PEESs, such as the instability risks of PLL caused by weak connections [74], voltage oscillation caused by the bandwidth of the current inner loop [140], AC current oscillation caused by improper filter parameters [141], and strong dynamic interactions between DC and AC equipment [142]. The stability mechanism can refer to that of REs and is not repeated in this study. However, if the DC/AC converter adopts active power control, the instability risks should be reanalyzed, while the analysis methods can remain the same as that of REs. For instance, negative impedance may occur when DC/AC converter uses active power control, resulting in oscillations as reported by [143]. In [144], feedforward power loops may imply high-gain instability in weak AC system clocks. The risk of PLL instability caused by weak connections under active power control has also been demonstrated in [70].

The impact of DC/AC converter on stability is not always negative considering various control strategies of DC/AC converters. In [145], an advanced control of PLL was proposed which reduces the negative effects on stability. In [146], the proposed control method can eliminate subsynchronous resonance. Therefore, a comprehensive evaluation should be conducted when applying modifications to fundamental control loops, avoiding potential risks caused by improper parameters or scenarios where the GFL control improves dynamic characteristics in one aspect but worsens others.

**Table 5**

Summary of instability risks when AC-PEES uses GFL control

| Outer control loop used | Instability loops | Causes and References |
|---|---|---|
| DC voltage control 1oop | PLL | Weak connections [74] |
| | Current inner loop | DC voltage oscillation [140] |
| | Filter loop | AC current oscillation [141] |
| | DC voltage outer loop | Strong dynamic interactions [142] |
| Active power control | Active power outer loop | Negative impedance [143] |
| | Active power outer loop | High-gain instability [144] |
| | PLL | Weak connections [70] |

### *5.1.2. Results of Dynamic Instability Risks when AC-PEES Uses GFM Control*

Considering the active supporting ability of PEESs, GFM control has gained much attention. Regarding GFM control, the traditional risks of GFL control may not be applicable, such as instability caused by weak connections. In [147], the ability of stable operation of GFM control under weak connections was verified, and in [148], the benefits of GFM control that can improve the damping of subsynchronous oscillations were demonstrated. However, although GFM control brings many advantages to power systems than GFL control, there are still some instability risks of GFM controls that should be taken into consideration [149]. For instance, the time-domain simulation method was used in [150] to analyze the impact of different control parameters of GFM control on dynamic characteristics, demonstrating the importance of design parameters. The modal analysis method was used in [151] to investigate the impact of dynamic interaction between PEESs and other equipment on stability, demonstrating risky conditions. In [147], the impedance-based analysis method was adopted, and instability risks are clarified in the frequency domain. In [152], a possibility of power and frequency oscillations was introduced, which can significantly reduce the stability of PEESs.

Although the stability analysis methods used for GFL and GFM controls are the same, the conclusions and instability risks differ [153]. The oscillation frequency of GFM control is normally lower than that of GFL control because the dynamics of GFM control are similar to that of synchronous generators. Therefore, there is a risk that GFM can participate or dominate low-frequency oscillations [154]. For instance, the studies in [154,155] demonstrated the possibility of GFM control participating in the low-frequency oscillations of SGs. The studies in [156] investigated the power oscillations dominated by multiple PEESs with GFM controls. Therefore, as the development of GFM control, whether a new type of oscillations may occur should be a concern.

### *5.1.3. Explanations on the Impact of Bidirectional Power Flow*

Similarly to DC-PEESs, power flow may also affect the stability results of grid-connected AC-PEESs. Here, two typical conditions are demonstrated. The first is when AC-PEESs adopt active power control, where the variation of output active power is approximately zero. Assuming the impedance of the AC-PEESs when outputting power is positive (as illustrated in Fig. 14(a) with the orange arrow), the impedance may change to be negative when the power flow is in the negative direction (as indicated by the blue arrow in Fig. 14(a)). This conclusion is very similar to that obtained in DC-PEESs, thus, detailed explanations are not repeated. In [157], similar conclusions were obtained by analyzing the effects of AC-PEESs with different power flow directions on the stability of connected AC power systems.

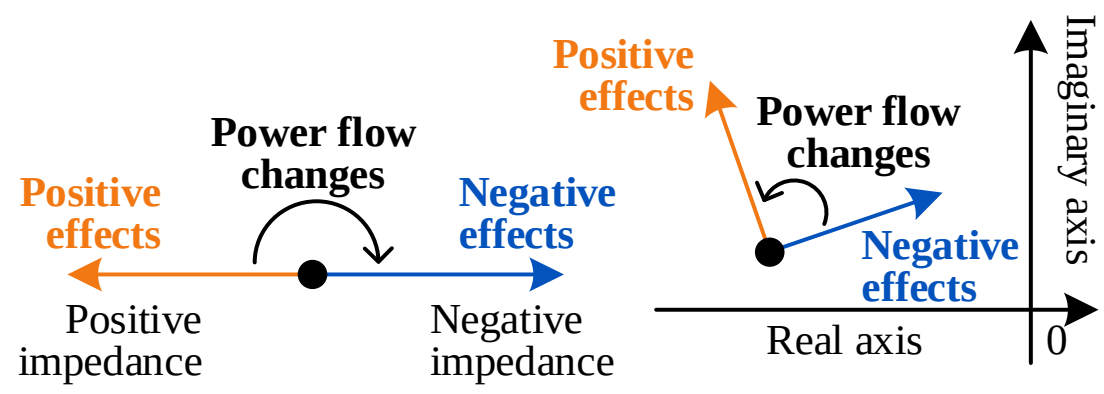

Fig 14. Illustration of the impact of bidirectional power flow of AC-PEESs.

The other condition involves changing the direction of interactions between any two AC-PEESs or the AC-PEES and another AC equipment. This conclusion can be derived using the open-loop modal resonance theory [158] and is explained as follows.

Assuming the transfer functions of two AC-PEESs as

$$\Delta y = \frac{P_1 K_1}{(s-\lambda)g_1(s)}\Delta x, \Delta x = \frac{P_2 K_2}{(s-\lambda)g_2(s)}\Delta y \tag{13}$$

where $\Delta x$ and $\Delta y$ are interconnected variables between two PEESs, and $\lambda$ is a common eigenvalue of them. $P_1$ and $P_2$ are the output power of the 1st and 2nd PEES, respectively, and the residual parts represent the remaining dynamics of the PEESs.

The eigenvalue variations caused by interactions between two PEESs can be solved from (13), as

$$\Delta\lambda = \pm\sqrt{P_1 P_2 K_1 K_2} / g(\lambda) \tag{14}$$

The result of (14) is illustrated in Fig. 14(b). It can be seen that the movement of eigenvalues rotates 90 degrees when the power flow directions of two PEESs change from the same to different, thus, the stability results are altered. However, the stability does not change in all cases when power flow reverses. If the instability is caused by factors that are unrelated to the power, the instability may not be affected or changed. For example, the instability caused by weak connections, where the major cause is the PLL [139], still occurs even if the power flow changes.

## *5.2. Issues and Solutions of Dynamic Analysis for Large-scale AC-PEESs*

As shown in Fig. 15, AC-PEESs can be connected in parallel or distributed ways. For parallel connections, all PEESs can be aggregated into one PEES [159]. Thus, stability analysis methods and results can be referenced from those of a grid-connected PEES. For example, in [159], the impact of the number of AC-PEESs on stability is analyzed using Bode plot by representing multiple PEESs as an equivalent PEES.

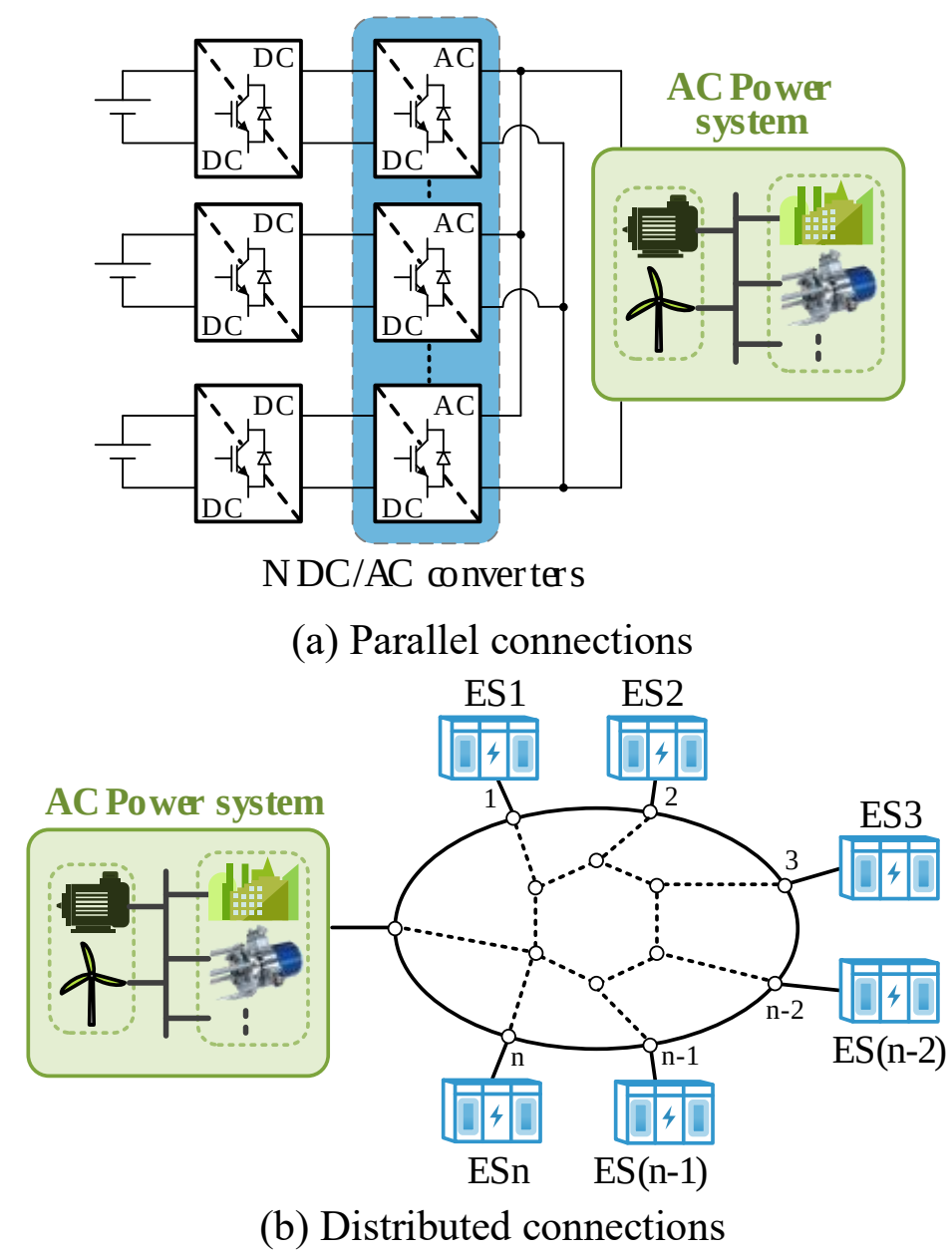

(a) Parallel connections

(b) Distributed connections

Fig 15. AC power systems connected with large-scale PEESs.

However, when multiple AC-PEESs connect to power systems in distributed ways, the interconnected power system is Multi-input and Multi-output (MIMO). For this case, it lacks mechanism analysis methods to reveal why this system can be unstable. Therefore, the stability of MIMO power system can only be evaluated by numerical calculation methods, such as modal analysis and Generalized Nyquist Criterion (GNC) based methods. For instance, the dynamic performance of the power system connected by multiple converters was assessed using the modal analysis [160]. In [161], the risk of low-frequency oscillations was identified for drop-controlled converters using GNC. However, the stability mechanism, that is, why these oscillations occur, remains unclear.

To apply the stability mechanism obtained by a single PEES to multiple PEESs, more studies attempt to use two PEESs as a tutorial example to replace multiple PEESs to obtain general rules of stability mechanism in multiple AC-PEESs. For instance, in [162], the N-1 converters were aggregated into one converter, and the interactions between N-1 converters and the other converters were simplified to be the interactions between two converters. In [163], any two of the grid-connected converters were selected, and their interactions were studied. It is indicated that strong interactions between any two converters can induce low-frequency oscillations.

**Table 6**

Summary of instability sources of PEESs

| Instability sources | DC-PEESs | | AC-PEESs | |
|---|---|---|---|---|
| | Single | Large scale | Single | Large scale |
| ESM parameters | [129,130] | \ | [108,138] | \ |
| Converter control parameters | [68,112,131] | \ | [70,144]<br>[147,150]<br>[154–156] | [142,151,152,158] |
| System parameters | [107] | [133–135] | [74,139,141,147] | [159–161] |
| Power flow | [67] | \ | [143,157,158] | \ |

Therefore, the stability mechanism of multiple AC-PEESs has been partially revealed for simplified scenarios. However, for large-scale PEESs without simplifications, the stability mechanism is a challenge to be revealed due to the lack of stability mechanism analysis methods for MIMO power systems. In addition to proposing such a method, another solution is dynamic reconfiguration as illustrated in Fig. 13. The details of instability sources of PEESs are summarized in Table 6.

## 6. Conclusions and Challenges

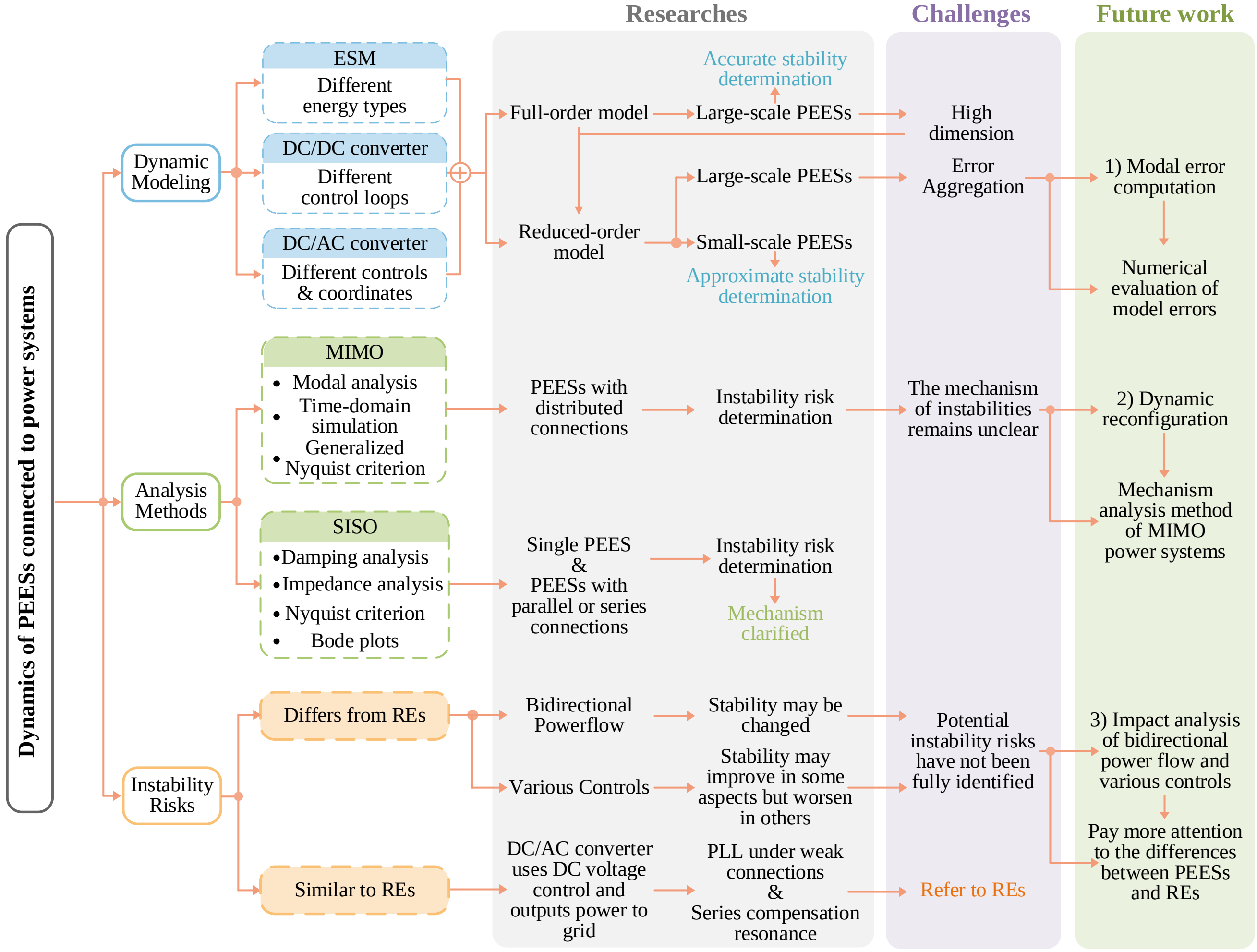

Fig 16. Summaries and future challenges.

In this study, the studies on dynamic analysis of connecting PEESs to power systems are reviewed, and instability risks are summarized, leading to the following three conclusions:

1) Comprehensive attention has been given to dynamic modeling of PEESs, covering various components such as ESMs, DC/DC converters, and DC/AC converters. However, in dynamic analysis, there tends to be a simplification of dynamics, with less focus on the errors introduced by these simplifications.
2) The integration of PEESs to both AC and DC power systems can lead to instabilities with different reasons. Notably, the popular GFM control, known for providing active support to power systems, also poses instability risks.
3) In simple scenarios where the power system is Single-input and Single-output (SISO), traditional stability mechanism analysis methods can be directly applied. This allows for the clear identification of how and why instabilities occur, including negative impedance/damping demonstrated by PEESs and strong dynamic interactions between PEESs and AC equipment. However, the instability mechanism of MIMO power systems has not been clear.

The challenges that should receive major attention fall into following three aspects: dynamic modeling, stability analysis methods, and potential risks.

1) The "error aggregation" caused by simplifications needs to be numerically clarified.
   It is crucial especially when considering large-scale integrations of PEESs. Two approaches are introduced: the first is calculating the impact of simplifications on the results, enhancing accuracy by considering their effects. The second proposes a modeling method that ensures both high accuracy and low complexity. In this study, one of the former methods is proposed to numerically evaluate the errors caused by model simplifications, improving the calculated results and enhancing overall accuracy.
2) The mechanism of instabilities caused by large-scale PEESs needs clarification.
   Traditional stability mechanism analysis methods can be directly applied to simple scenarios where the system can be established as a SISO power system. However, for MIMO power systems, two approaches are proposed to reveal the stability mechanism. The first involves reconfiguring the dynamics of the power systems into identical and distinct parts. Analyzing the results from these aspects can help clarify the causes of instabilities in PEESs. The second approach introduces a general method for analyzing the stability mechanism in MIMO power systems, which still requires further investigation. In this study, the proposed dynamic reconfiguration method can also be used to quickly calculate the dominant oscillation mode, enabling rapid numerical decisions in stability analysis.

3) Applying risks of REs directly to PEESs may overlook specific risks of PEESs, which should be reevaluated. Two scenarios requiring major concern beyond RE risks are identified: firstly, the various control strategies for PEESs proposed by studies may enhance dynamic characteristics but could also introduce potential instability risks that need careful examination. Secondly, bidirectional power flow caused by PEESs, an area receiving less attention currently, could lead to new instabilities, differing significantly from traditional scenarios where power flow is unidirectional from the REs side to the load centers. Additionally, two typical examples are provided to illustrate the impact of bidirectional power flow on impedance and dynamic interaction. These scenarios should be carefully considered, and further exploration is needed for other situations.

**Declaration of competing interest**

The authors declare that they have no known competing financial interests or personal relationships that could have appeared to influence the work reported in this study.

**Acknowledgments**

This work is supported by the National Key Research and Development Program of China under Grant 2023YFB2405900. We also acknowledge the support from the Hong Kong Scholars Program.